\documentclass[twocolumn,showpacs,preprintnumbers,amsmath,amssymb]{revtex4}

\usepackage{graphicx}
\usepackage[pdftex]{hyperref}
\usepackage{dcolumn}
\usepackage{bm}

\begin{document}
\title{Consequences of the Production of Very Massive Magnetically Charged
Leptons Early in the Universe and Their Decays to a New Set of Extremely Massive Neutrinos}
\author{Sherman Frankel}
\email{frankel@physics.upenn.edu}
\affiliation{%
University of Pennsylvania, Department of Physics \& Astronomy\\
209 S. 33rd Street, Philadelphia, PA 19104}%

\date{\today}

\begin{abstract}

We examine the production and decay of extremely heavy
magnetically charged leptons,($\tau_g$ and $\mu_g$), to their own very heavy
$\mu_g$ and $e_g$ plus their own new species of neutrinos, $\nu_g$ and
$\bar{\nu}_g$, at some time early in the universe which could be present in
space and, attracted gravitationally towards and passing through
astronomical objects, annihilated with each other to produce large
numbers of photons. Further, we describe the possibility of presently
detecting the bursts of such photons, of three different total
energies, in the seas or oceans on earth.
\end{abstract}

\maketitle

 A century has passed since the idea of the existence of magnetic
 charges was introduced by modifying the equations of electromagnetism
 so that $div B$ was not equal to zero but replaced by $div B = \rho_g$ in
 analogy with $div E = \rho_e$. Indeed in 1904 it was shown \cite{thomson}
 that for
 a monopole $g$ at rest, separated from an ordinary charge $e$ at rest,
 that there was an angular momentum in the field given by $\vec l = ge \;
 \vec r_1$. In 1931 it was this angular momentum that P.A M. Dirac
 quantized \cite{dirac} which then required that, while the fine structure
 constant, $e^2/\hbar c$, was 1/137, the similar constant, $g^2/\hbar c$, was
 huge and equal at least to 137.  In 1976 \cite{scalar} it was realized
 that the
 symmetry requirement that electromagnetism must conserve parity (P)
 and be time reversal invariant (T) allowed one to derive the above
 results without any use of Maxwell's Equations.

Now, decades later, we know of the existence of the three
 electrically charged leptons tau, mu, and e. Indeed, because the fine
 structure constant is so small, one can actually calculate the exact
 masses of electrically charged leptons, tau and mu.  Unfortunately,
 because $g^2/\hbar c$ is so large, one cannot calculate the magnetic
 lepton masses.  For many decades, as accelerator energies have
 increased, searches for ``monopoles'', as they are often called, have
 continued and most recently this journal has again reported the
 failure to find them at the present high energies available in the
 Fermilab collider \cite{fermilab}.  

There does not appear
 to be any reason that
 the hierarchy of magnetically charged particles should be any
 different from the hierarchy of the known electrically charged
 particles, but this note will not address the interesting questions
 of magnetically charged $W$'s, other than to remark that they must
 exist.

The first prediction is that the magnetic leptons, the $\tau_g$ and $\mu_g$,
will decay to their $e_g$ by the same reaction found for ordinary
electrically charged tau's and mu's that decay to the neutrinos that
have been observed. However, this will result in the creation of a new
set of magnetic neutrinos much heavier than the neutrinos from
ordinary tau and mu decays which we now know to have extremely small
masses, not equal to zero.  The ordinary (electric origin) neutrinos
that have been measured till now are often called ``massive neutrinos''
and a complete review of their properties can be found in reference
\cite{habib}. Perhaps, to stress the distinction, they should be called light
neutrinos.)

We now turn to the interesting properties of the new neutrinos that
are involved in the decays of magnetically charged tau's and mu's:

\begin{enumerate}
\item They do not interact at all with the protons, neutrons and
electrons that make up our universe. Only ordinary neutrinos can do
so. (This would not be the case in a universe made up of anti-matter.)
\item They have mass so they are attracted to massive bodies like the
earth, but they simply pass on through, making no interactions with
anything. 
\item  However, there are occasions when a magnetic neutrino
comes within range of a magnetic antineutrino. The result would be a
burst of photons with three different energies depending on whether
the neutrinos are associated with the different mass magnetic leptons,
tau, mu, or e. 
\item It is perhaps worth noting that the same Feynman
diagram used to describe the annihilation of ordinary neutrinos and
their antineutrinos into photons would be used but with a new
magnetically charged $W$ as the intermediate state.
\end{enumerate}

The bursts of photons coming from the magnetic origin proposed
neutrino annihilations should be examined by astrophysicists since
they should have affected the expansion of the universe.

We next examine the possibility of an experiment on earth that might
possibly actually measure the photons from the annihilations of
magnetic origin neutrinos.  Experiments in mines have the property
that the neutrinos would be attracted into the mine walls and, if they
annihilate within, the photons would be absorbed and not enter the
detectors.  However experiments in seas and oceans would not have this
problem. In particular there is the experiment called NESTOR that has
its laboratory on the shores of the Ionian Sea at Pylos. Having a
large number of layers of detectors in large rings that go miles deep
into the Sea, and not surrounded by walls, it would be a place to
search for the photon bursts since it is designed, among other
things, to detect photons \footnote{Leonidas Resvanis is Professor of
Physics at the University of Athens and director of the NESTOR
Institute for Astroparticle Physics of the National Athens
Observatory.}.

Searches for monopoles and massive exotic particles have been of great
interest and the reader might want to browse the reports of various
collaborations: The MACRO collaboration, ``Search for GUT Monopoles
and massive Exotic Particles''; Antares Collaboration, ``Neutrino
Telescopes as Magnetic Monopole Detectors''; Inst. For Nuclear Research
of RAS, Moscow, ``Detection of Relativistic Magnetic Monopoles''.

\begin{acknowledgments}I should like to thank Eugene Beier, Bhuvnesh Jain,
Jeffrey Klein, Ken Lande, Burt Ovrut, Gino Segr\'{e}, Paul Steinhardt
 and other of my Penn colleagues
for useful discussions of magnetic charge over the years.
\end{acknowledgments}

\bibliography{magnetic_monopoles}

\end{document}